\documentclass[aps,pra,preprint]{revtex4}
\usepackage{amsmath}
\usepackage{amssymb}
\usepackage{graphicx}

\begin{document}

\title{Induced magnetism in transition metal intercalated graphitic systems}

\author{T. P. Kaloni}
\author{M. Upadhyay Kahaly}
\email{mousumi.upadhyaykahaly@kaust.edu.sa}
\author{U. Schwingenschl\"ogl}
\email{udo.schwingenschlogl@kaust.edu.sa}

\affiliation{PSE Division, KAUST, Thuwal 23955-6900, Kingdom of Saudi Arabia }
\begin{abstract}
We investigate the structure, chemical bonding, electronic properties, and magnetic behavior of 
a three-dimensional graphitic network in aba and aaa stacking with intercalated transition
metal atoms (Mn, Fe, Co, Ni, and Cu). Using density functional theory, we find induced
spin-polarization of the C atoms both when the graphene sheets are aba stacked (forming graphite)
and aaa stacked (resembling bi-layer graphene). The magnetic moment induced by Mn, Fe, and 
Co turns out to vary from 1.38 $\mu_B$ to 4.10 $\mu_B$, whereas intercalation of Ni and Cu 
does not lead to a magnetic state. The selective induction of spin-polarization can be utilized in 
spintronic and nanoelectronic applications.  
\end{abstract}
\maketitle

\section{Introduction}

The rich variety of the physical and chemical properties of nanoscale materials is used in a wide range 
of advanced technological applications. Graphene, a single layer of graphite with C atoms tightly 
packed in a honeycomb lattice \cite{Novoselov}, is one of the most promising low-dimensional nanomaterials,
attracting immense interest of both experimentalists and theoreticians \cite{tan,son} because of its 
two-dimensional (2D) structure and unique properties. In both pristine graphene and graphite no magnetic 
ordering is expected, by experiment and theory. However, there are various experimental methods to
induce magnetism: hybridization between the C $p_z$ and Ni $d$ states \cite{Weser},
ion implantation \cite{Palacio}, proton irradiation of highly oriented pyrolytic graphite \cite{Esquinazi,Esquinazi1},
and point defects \cite{Flipse1}. Ferromagnetism can be induced in multilayered graphene by randomly
removing single C atoms \cite{Ugeda}. Theoretically, spin-polarization has been confirmed for various C defects 
\cite{Lehtinen,Ricardo2,Zhang,Lisenkov,Ricardo1,oleg,Kaloni}, adsorption of molecular oxygen and hydrogen on graphene 
\cite{dai}, and substitutional Mn doping of graphene \cite{wu}.

Inducing spin-polarization in graphene by doping transition metal (TM) atoms \cite{uchoa,uchoa1,Foster} is important 
because it can lead to scattering \cite{Guinea} and modify the electronic states locally, which is required for graphene-based 
electronics and Kondo physics. Often, very simplistic assumptions about the effects of the TM atoms are not valid, but the properties 
depend strongly on the host \cite{lany}. Furthermore, precise knowledge of the TM-C interaction is important for understanding
carbon nanotube growth \cite{yazyev}, fuel cells \cite{che}, and the role of implanted magnetic atoms, such as
Fe, for the magnetic order \cite{sielemann}. For these reasons it is surprising that TM intercalation in a three-dimensional 
(3D) graphitic network has not been studied with respect to the modifications of the interlayer interaction.
In this paper, we will present first-principles results on the spin-polarization induced by Mn, Fe, Co, Ni, and Cu atoms intercalated 
between adjacent C layers. We study two types of stacking: aaa, which results in multilayer graphene, and aba, which
results in bulk graphite. The different $d$ valences of the TM atoms result in a range of induced magnetic moments 
(magnitude and spatial distribution) as needed for spintronic applications.

\section{Methods}

Calculations based on density functional theory \cite{kohn} are carried out using the generalized gradient approximation (GGA) 
and the plane wave Quantum-ESPRESSO code \cite{paolo}, which already has been used successfully for describing graphene 
\cite{Cohen,car,cheng,cheng1,kaloni,das}. A strong corrugation of the crystal potential is present in layered materials 
in the direction perpendicular to the atomic planes. Consequently, the local density approximation is not valid. 
Even though the van-der-Waals forces are neglected in both the local density approximation and the GGA, the 
GGA leads to a reasonable matching of the interlayer distances with experiments and is adequate for systems with an inhomogeneous
charge density \cite{blaha,dantas,kar}. We apply Perdew-Burke-Ernzerhof pseudopotentials \cite{pbe} 
and include an onsite Coulomb interaction correction. In our calculations scalar relativistic effects are included, 
while spin-orbit coupling is not taken into account as it is negligible in the present systems. The value of the onsite interaction 
$U$ is varied between 0 and 8 eV. Our results show that it is essential to use a non-zero $U$ for the $d$ electrons of the TM atoms.
Previous findings suggest that $U$ (which is mainly an atomic property) ranges from 1 to 5 eV for TM atoms \cite{qi,bocquet}. 
For $U\geq2$ eV our results are qualitatively similar with small variations in bond lengths and magnetic moments ($<5$\%).
Hence, we will discuss data for $U=4$ eV in the following, unless specifically mentioned. A
large $U$ of 8 eV leads to a reduced cohesive energy and to overestimated magnetic moments, especially in the case of Mn. 
All calculations are performed with a plane wave cutoff energy of 408 eV. We use a Monkhorst-Pack \cite{Monk} 
16$\times$16$\times$4 k-mesh for the calculation of the band structure and 
density of states (DOS) \cite{wu,udo}.

We find that a 3$\times$3 supercell is large enough to avoid drawbacks of the periodic 
boundary conditions. Our supercell contains 36 C atoms and one dopant in all cases, with a unit cell length of 7.45 \AA\ 
in the $ab$-plane. The length of the $c$-axis is relaxed ($\sim$7 \AA). 
It was suggested that the energy cost associated with the formation of two separated magnetic moments is given by 
the RKKY coupling between two magnetic impurities embedded in a metal, which decays slowly with the 
separation of the impurities in the case of a low-dimensional system. Consequently, inclusion of this effect would demand a huge unit
cell with several magnetic impurities, exceeding reasonable computational resources.
We fully relax the positions of all the atoms, including the dopants, in order to obtain the minimum energy configuration. 
We continue the optimization until an energy convergence of 10$^{-7}$ eV per supercell and a force convergence 
of 0.04 eV/\AA\ is reached. To avoid trapping of TM atoms in local energy minima 
we start have studied different starting configurations where the TM atom is located 
above either the bridge site (center of a C$-$C bond), or the hollow site (center of a C hexagon)
or the top site (top of a C atom). The final lowest energy structures are tabulated in Fig.\ 1.   

\section{Structures and stability}

Our data suggest that the TM atoms prefer being accommodated in different sites when intercalated in aaa and aba hosts.
In aaa stacking Fe, Co, Ni, and Cu atoms tend to move towards the hollow site, while Mn atoms 
(smallest atomic number $Z$) prefer to remain next to a C atom and form an almost linear C$-$Mn$-$C bond. 
However, in aba stacking Mn, Fe, Co, and Ni atoms favor the hollow site of one C layer, which is the
top site of the next C layer, see Fig.\ 1. Cu atoms (highest $Z$) are slightly off-centered
as compared to the other atomic species.

The observed structural variations can be explained by the crystal field splittings which apply to the $d$ 
orbitals of the different TM atoms. If a spatially symmetric negative charge is placed around a particular TM atom, the $d$ 
orbitals remain degenerate. However, being itself negatively charged the TM atom experiences repulsion 
and the orbital energies are raised. As in our case the field results from electrons in the
C layers on both sides of the TM atom, the charge distribution is not spherical and the 
degeneracy of the $d$ orbitals is lifted. The effect is stronger for atoms with partially
filled $d$ orbitals (Fe, Co, Ni, and Cu), which thus 
tend to distort away from the trigonal symmetric position and form a tetrahedral-like
structure with the C atoms. This is clearly visible in the left column of Fig.\ 1. 
The average C$-$TM$-$C bond angles for Mn, Fe, Co, Ni, and Cu turn out to be
180$^{\circ}$, 143$^{\circ}$, 151$^{\circ}$, 144$^{\circ}$, and 155$^{\circ}$ in aaa stacking and 
143$^{\circ}$, 139$^{\circ}$, 141$^{\circ}$, 141$^{\circ}$, and 164$^{\circ}$ in aba stacking. 
The Mn atom, being exactly half 
filled, experiences only a weak effect and retains the trigonal symmetry, forming a linear C$-$Mn$-$C bond. 
In the aba stacking there is additional negative charge due to the C $p_z$ orbitals. 
Here, the shift of neighboring C layers within the $ab$-plane offers the TM atoms a tetrahedral-like geometry. 
However, the unpaired electron of the Cu atom results in some off-centering. 

We address the cohesive energy ($E_{coh}$) in Table I in order to describe the stability of the different structures.
The cohesive energy per atom is defined as
\begin{equation}
E_{coh} = \frac{(E_{cell} - \sum E_{atom}-E_d)}{n},
\end{equation}
where $E_{cell}$ is the (spin-polarized) total energy of a cell containing $n$ atoms. $E_{atom}$ and $E_d$ are the 
total energies of isolated C and dopant atoms, respectively. They are calculated by placing a single atom 
in a sufficiently large cubic unit cell of $\sim$11 \AA\ lateral length, in order to avoid interaction with the periodic image.

The calculated magnetic moment, smallest distance between the TM atom and its nearest neighbor C atom,
cohesive energy, change of the in-plane C$-$C bond lengths, and estimated buckling are summarized in Table I.
In aaa stacking the TM$-$C bond length for Mn is 2.00 \AA, while it varies for Fe, Co, Ni, and Cu in the 
ranges 2.03 to 2.26 \AA, 2.01 to 2.04 \AA, 1.92 to 2.18 \AA, and 1.97 to 1.98 \AA, respectively.  
In aba stacking the TM$-$C bond lengths for Mn vary in the range 2.00 to 2.39 \AA, while for Fe, Co, Ni, and Cu we 
obtain 2.02 to 2.19 \AA, 2.02 to 2.30 \AA, 1.96 to 2.22 \AA, and 1.97 to 2.19 \AA, respectively.
Both the bond lengths and bonding energies in Table I indicate that the bonding between the TM atoms 
and the C layers is covalent.
A significant buckling is observed in the C layers due to the repulsion of the TM charge, amounting 
to 0.185 \AA, 0.093 \AA, 0.168 \AA, 0.107 \AA, and 0.098 \AA\ (aaa stacking) and 0.118 \AA, 0.015 \AA, 
0.079 \AA, 0.044 \AA, and 0.098 \AA\ (aba stacking) for Mn, Fe, Co, Ni, and Cu, respectively. The buckling
is calculated as the difference between the $z$-coordinates of the highest and lowest C atoms in a particular layer.
The in-plane C$-$C bond lengths range from 1.43 \AA\ to 1.45 \AA\ in all structures, showing that the sp$^2$-hybridized nature of the in-plane bonding is retained. 

\section{Electronic structure and magnetism}

E$_{coh}$ and the bonding strength decrease monotonically with increasing onsite interaction $U$ 
(with a slight exception for Fe), see the top panels of Fig.\ 2. This has a clear but limited effect on the 
magnetic properties induced by the dopants, as is evident from the bottom panels of Fig.\ 2.
We note that Mn shows a saturation of the magnetization with increasing $U$, consistent with Ref.\ \cite{wu}, 
whereas for $U$ = 8 eV the magnetization trails off for Fe and Co. A growing value of $U$ will affect
more prominently the partially filled $d$ systems Fe and Co. A choice of $U=4$ eV is reasonable for our purpose.  

In contrast to Mn doped graphene \cite{wu}, Mn intercalated graphite develops metallic states, see Fig.\ 3. 
The spin-polarized DOS for $U$ = 4 eV shows that the presence of the TM enhances the DOS at the Fermi level. Both stacking schemes
induce spin-polarization for Mn, Fe, and Co intercalation, amounting to 4.10 $\mu_B$, 3.70 $\mu_B$, and 1.97 $\mu_B$ (aaa stacking) 
and 3.80 $\mu_B$, 2.06 $\mu_B$, and 1.83 $\mu_B$ (aba stacking). Ni and Cu intercalation does not result in a magnetic state for both 
aaa and aba stacking. For all dopants the induced magnetic moment differs in aaa and aba stacking, due to different average C$-$TM bond 
lengths. The smaller this length, the larger is the magnetization.
The spin density induced by Mn, Fe, and Co is addressed in the 3rd and 5th column of Fig.\ 1.
The maps show localized magnetic moments on the C atoms, which reveal
characteristic patterns of parallel and antiparallel orientations. When the total magnetic
moment decreases along the series Mn-Fe-Co, these localized moments reduce in magnitude.
However, simultaneously a weak spin density develops in the interstitial region between
the atomic sites which shows a distinct circular wavy shape. This delocalized spin-polarization
even dominates in the case of Co intercalation. 

A L\"owdin charge analysis shows that the Mn atom loses almost 0.5 electrons when it is intercalated, with $d$ orbital occupations of
$\sim$5.5 electrons in aaa and $\sim$5.6 electrons in aba stacking. Due to the trigonal-like crystal field around the Mn atom, the 
fivefold degeneracy of the $d$ orbitals is lifted and the $d_{3z^2-r^2}$ orbital becomes almost fully occupied (in a comparatively weak crystal field).
Hence, the 5.5 electrons in aaa stacking yield local magnetic moments of 3.5 $\mu_B$ to 4 $\mu_B$, while in aba stacking the 5.6 electrons
yield a slightly smaller moment. We find that Fe loses almost 0.4 electrons in aaa and 0.3 electrons 
in aba stacking. The occupations of the $d$ bands are 4.8 and 4.7, respectively. A stronger crystal field results in partial occupations of 
the  $d_{3z^2-r^2}$, $d_{xz}$, and $d_{yz}$ orbitals and total magnetic moments of 3.70 $\mu_B$ and 3.06 $\mu_B$ for aaa and aba stacking.  
A very strong crystal field splitting for Cu and Ni results in partial filling of all $d$ orbitals 
and a prominent out-of-plane bonding of the $d$ orbitals with the C $p_z$ orbitals, resulting in zero magnetic moment.                                                                                   

Analysis of the DOS reveals that for Mn intercalation and aba stacking the states below the
Fermi level reveal a strong contribution of the C $p$ orbitals, while the states just above the
Fermi energy are dominated by the Mn $d$ orbitals. For both aaa and aba stacking, the DOS
analysis suggests that the Mn atom is in a Mn$^{2-\delta}$ state, where $\delta$ is a small
positive number. The partial $s$, $p$, and $d$ occupations reflect the non-ionic character of
the bonding. In the case of Fe intercalation, for both aaa and aba stacking, the states below
the Fermi level are entirely due to the C $p$ orbitals, while the lowest unoccupied states are
a mixture of C $p$ and Fe $d$ orbitals. In case of Ni and Cu intercalation, we find that the
$d$ orbitals do not contribute to the electronic states around the Fermi level, which mainly
originate from C $p$ and TM $s$ orbitals. This explains the non-magnetic behavior of Ni and Cu.
In general a mixing of C $p$ and TM $d$ states points to a partially covalent character
of the TM$-$C bond. 

\section{conclusion}
In view of the possibility to realize a Kondo system by chemisorption of metal atoms on graphene \cite{Kondo},
we have addressed the behavior of the TM atoms Mn, Fe, Co, Ni, and Cu when they are
intercalated in a 3D graphitic network with aaa and aba stacking.
While TM adatoms on pristine graphene are reported to have bonding energies of some 0.2 eV to 1.5 eV \cite{sevincli,chan},
we obtain bonding energies between 1.10 eV and 4.23 eV for the systems under investigation. 
The electronic structure of the graphitic systems is modified significantly after intercalation 
due to hybridization of the C $p_z$ orbitals with the TM $d$ orbitals. It turns out that Mn, Fe, and Co
induce spin-polarization in both stacking configurations, whereas Ni and Cu result in metallic systems with zero magnetic moment. 
While graphene is characterized by a high mobility of incorporated TM atoms, the strong preferential
bonding of intercalated TM atoms in our case paves the way to graphitic Kondo systems.

\begin{acknowledgments}
We thank KAUST research computing for providing the computational resources used for this study.
\end{acknowledgments}

\begin{figure*}[p]
\begin{tabular}{|c|c|c|c|c|c|}
\hline 
System & aaa  &Spin density& aba & Spin density& Colour code \tabularnewline
\hline
Mn &\includegraphics[scale=0.3]{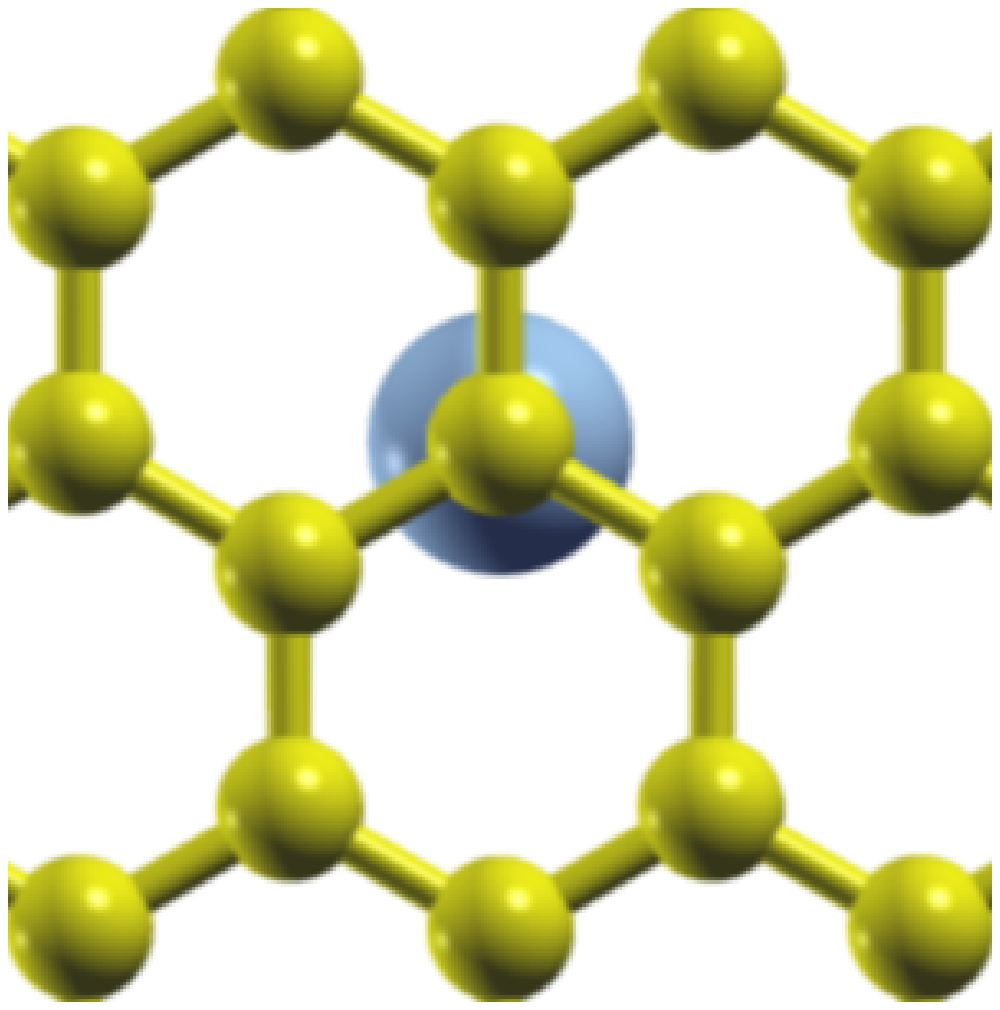}&\includegraphics[scale=0.3]{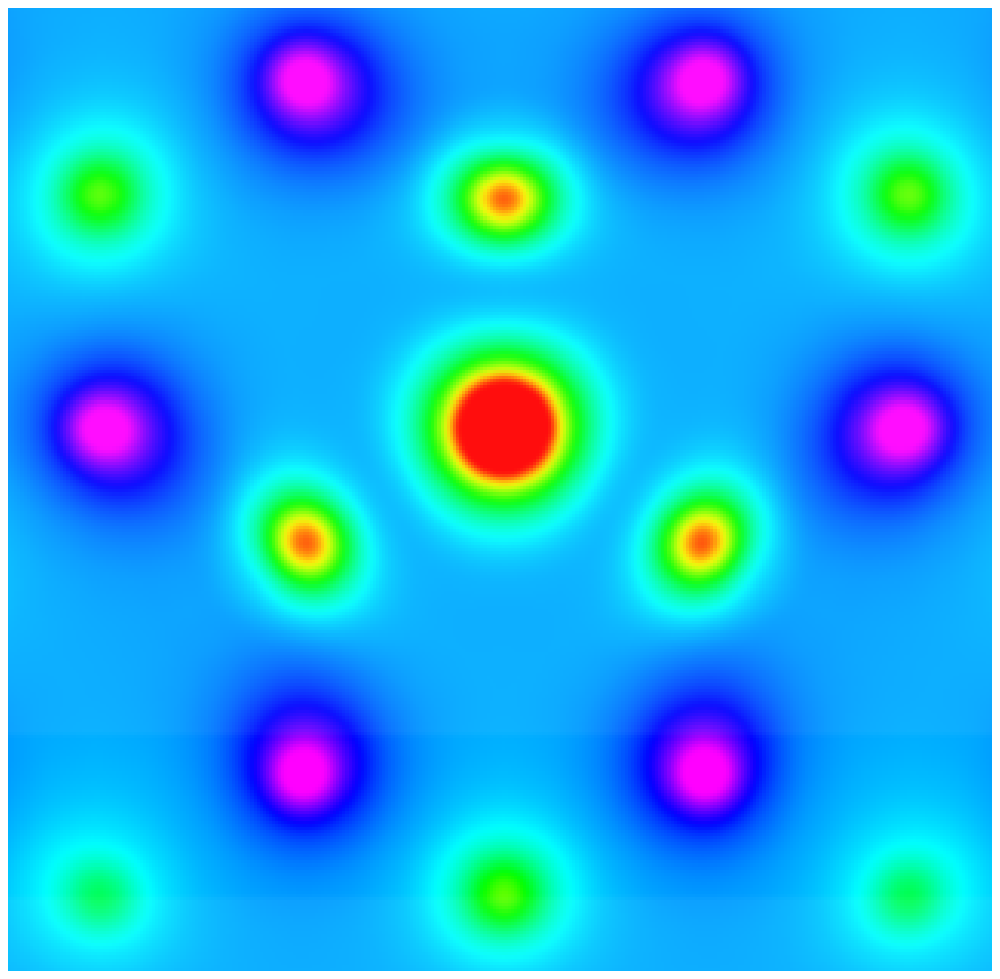}&\includegraphics[scale=0.3]{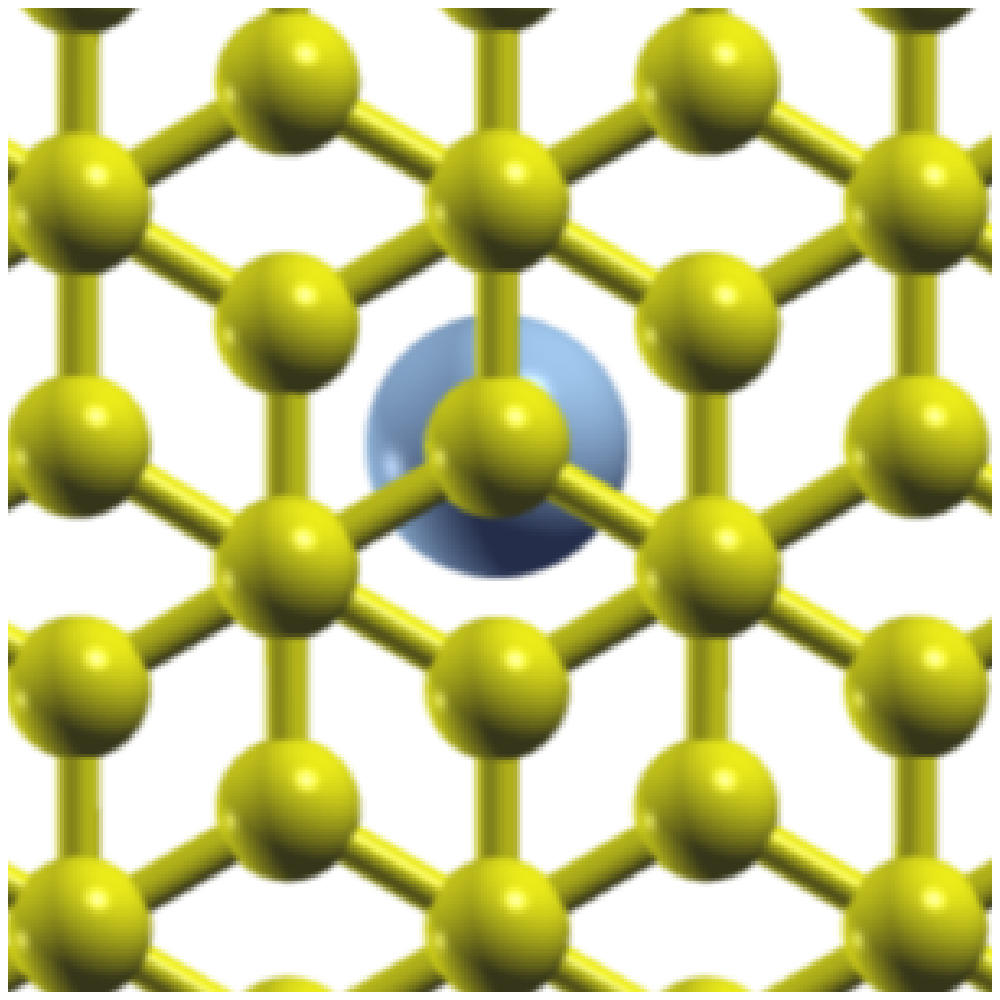}&\includegraphics[scale=0.3]{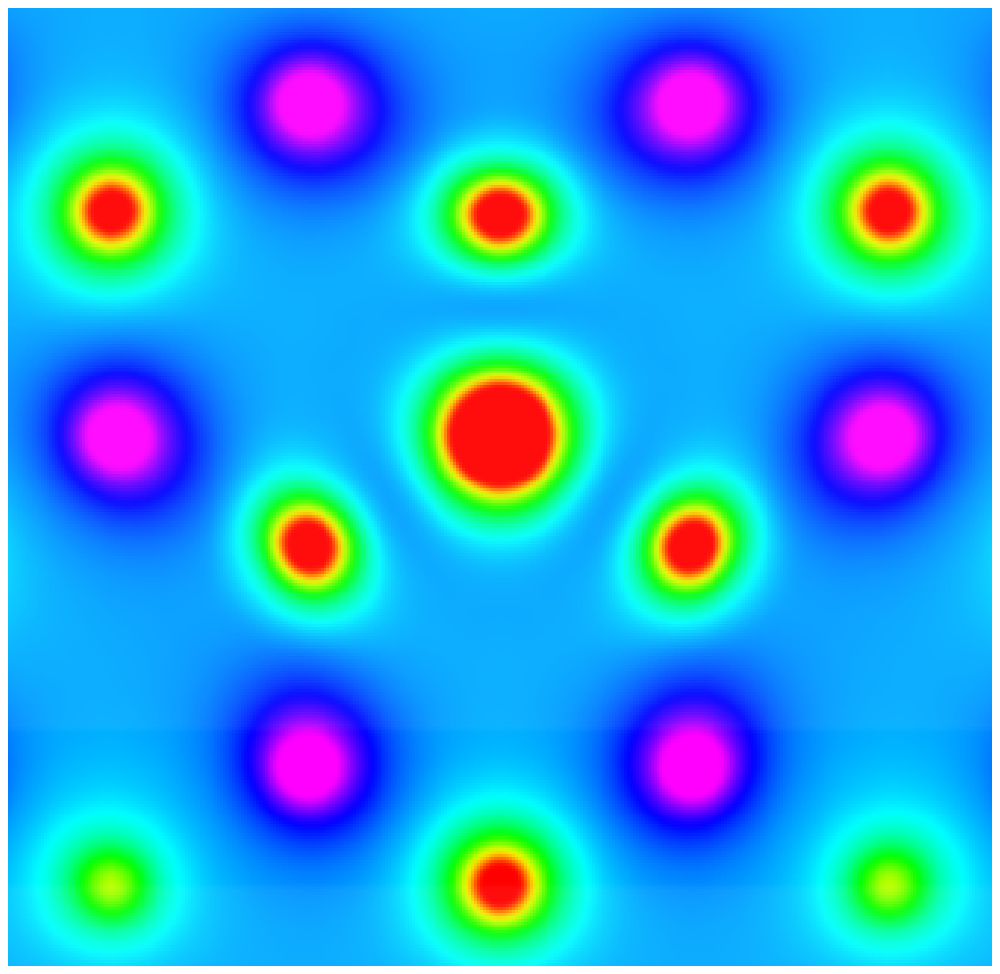}&\includegraphics[scale=0.45]{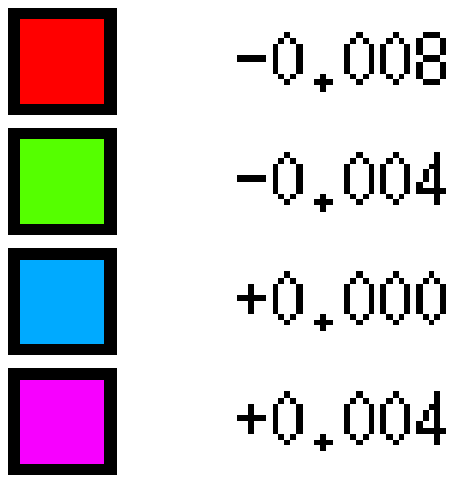}\tabularnewline
\hline 
Fe& \includegraphics[scale=0.3]{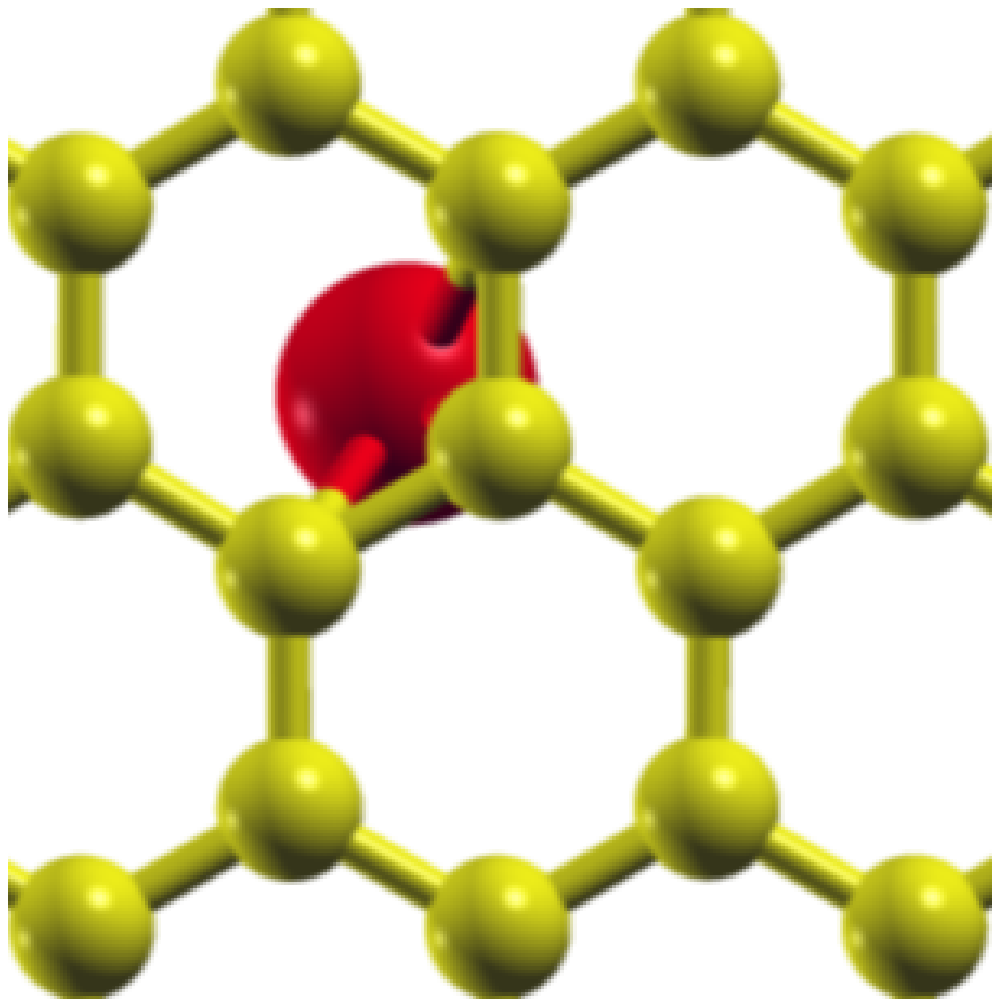}&\includegraphics[scale=0.3]{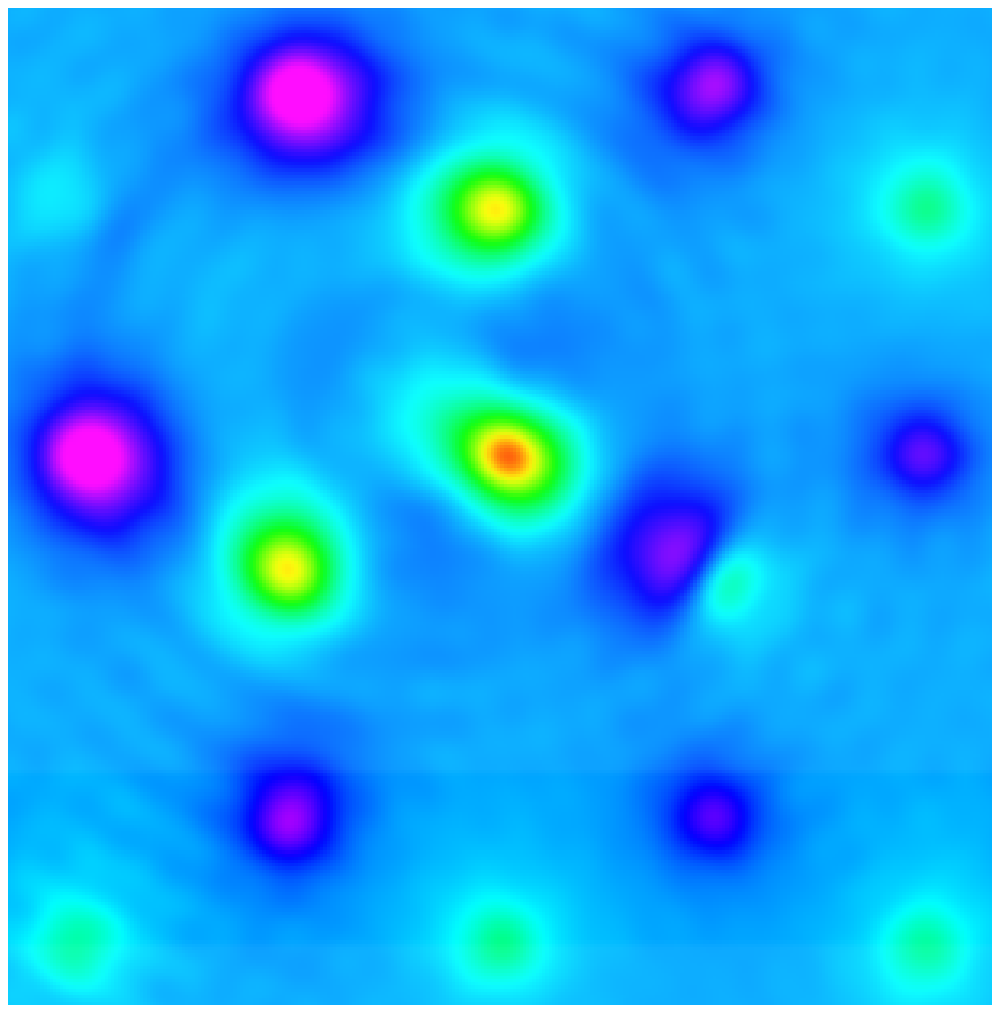}&\includegraphics[scale=0.3]{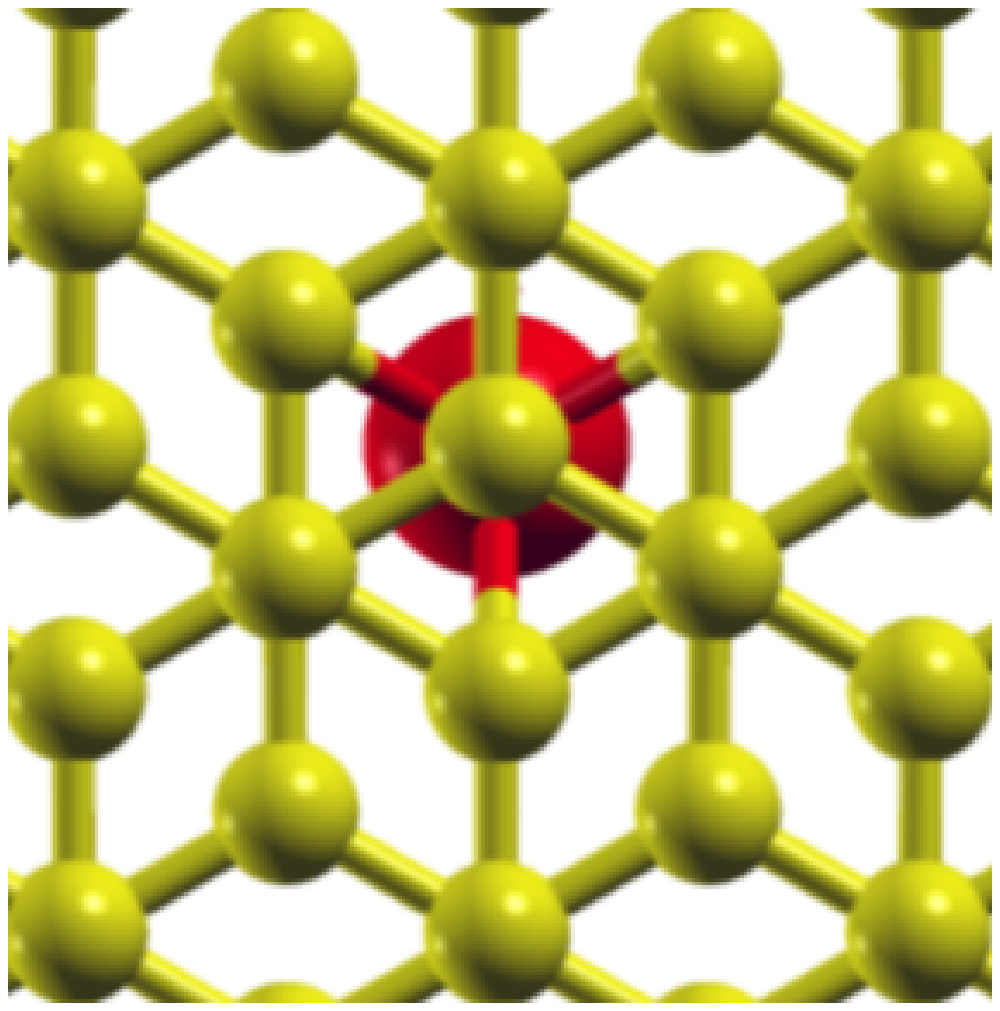}&\includegraphics[scale=0.3]{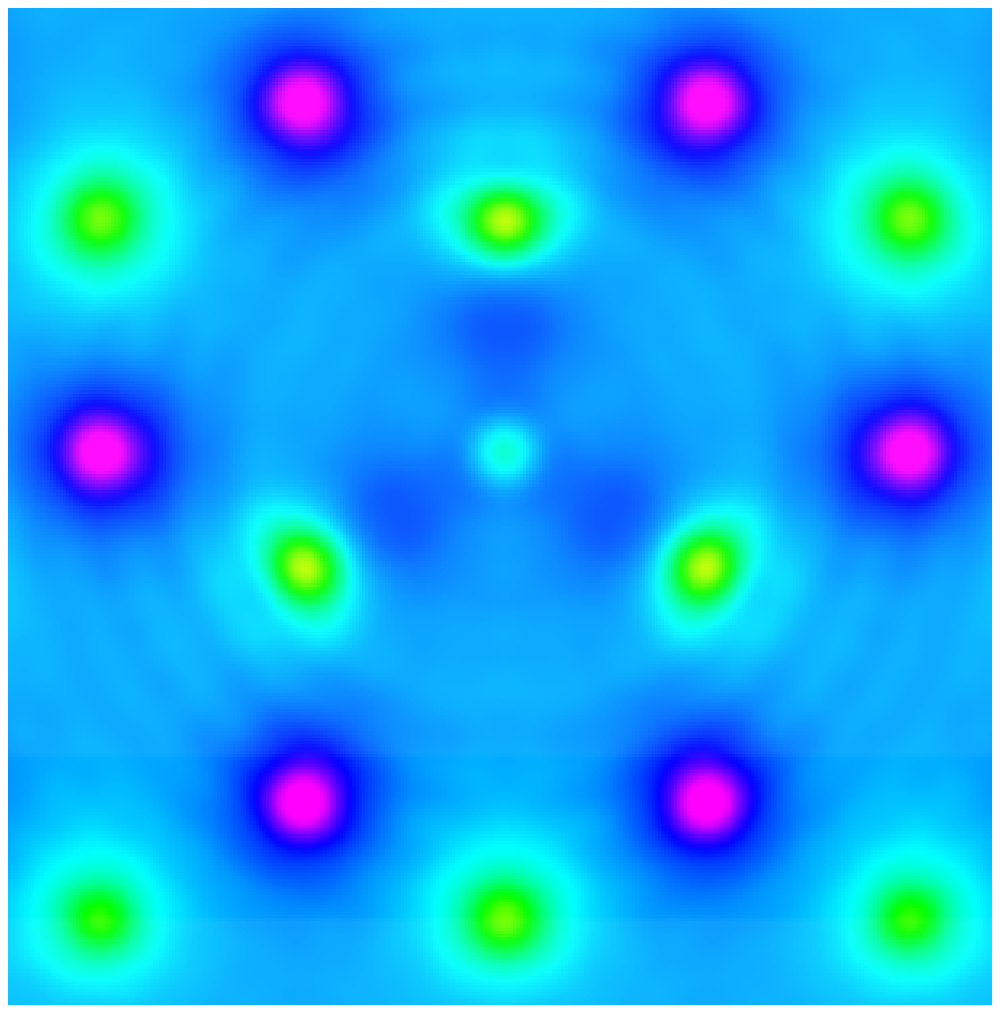}&\includegraphics[scale=0.45]{t}\tabularnewline
\hline 
Co& \includegraphics[scale=0.3]{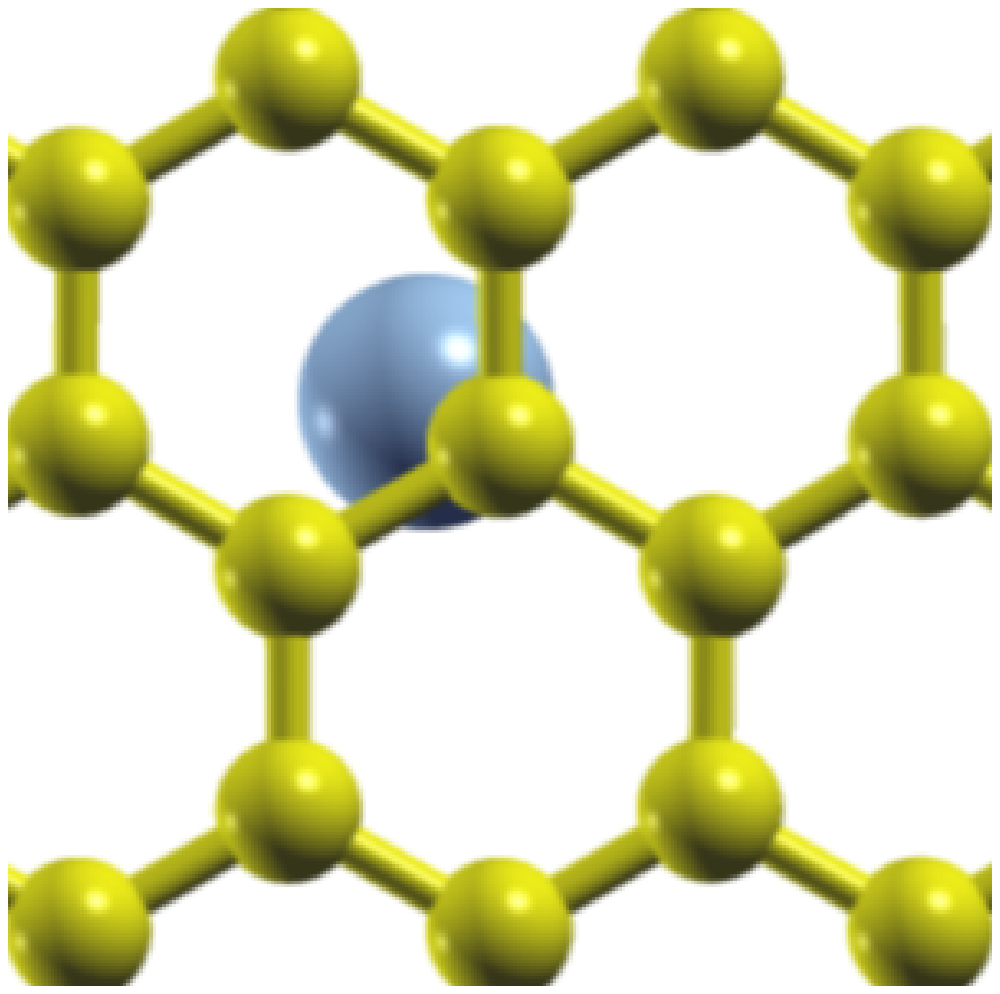}&\includegraphics[scale=0.3]{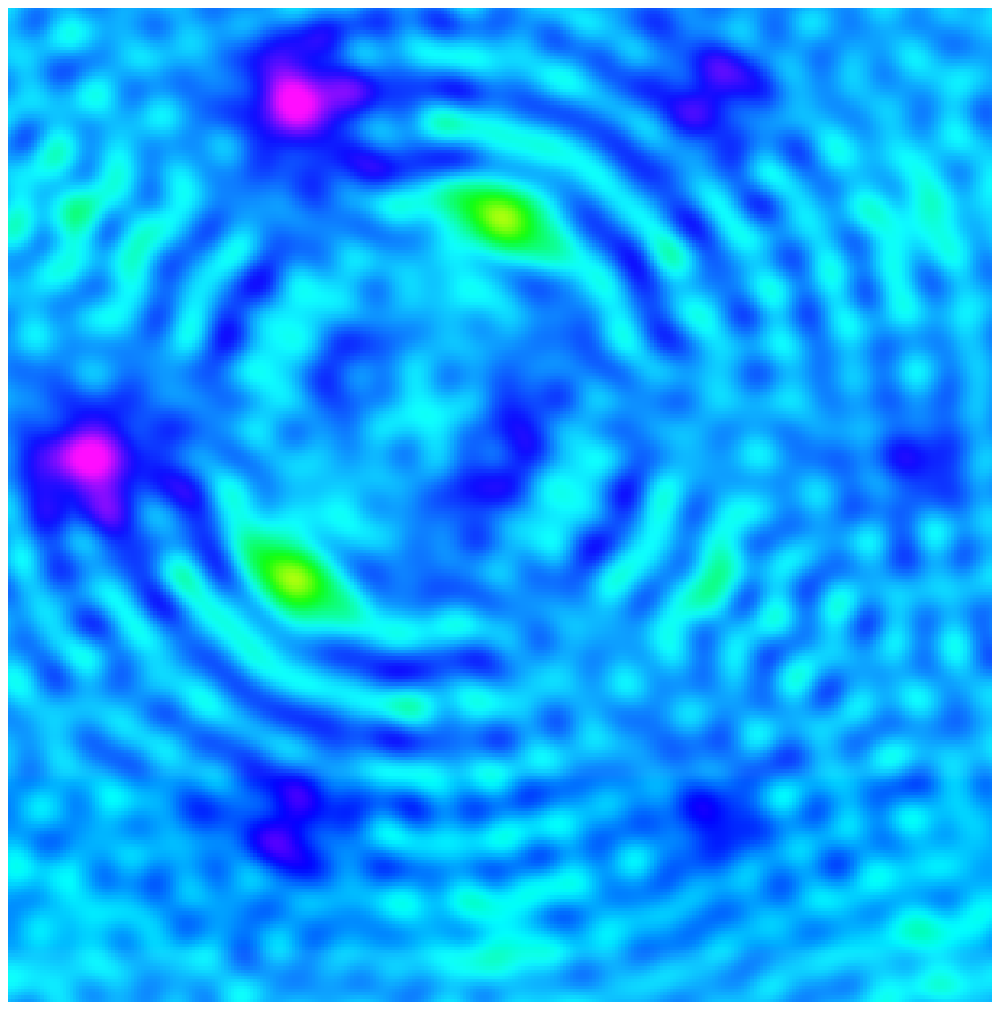}&\includegraphics[scale=0.3]{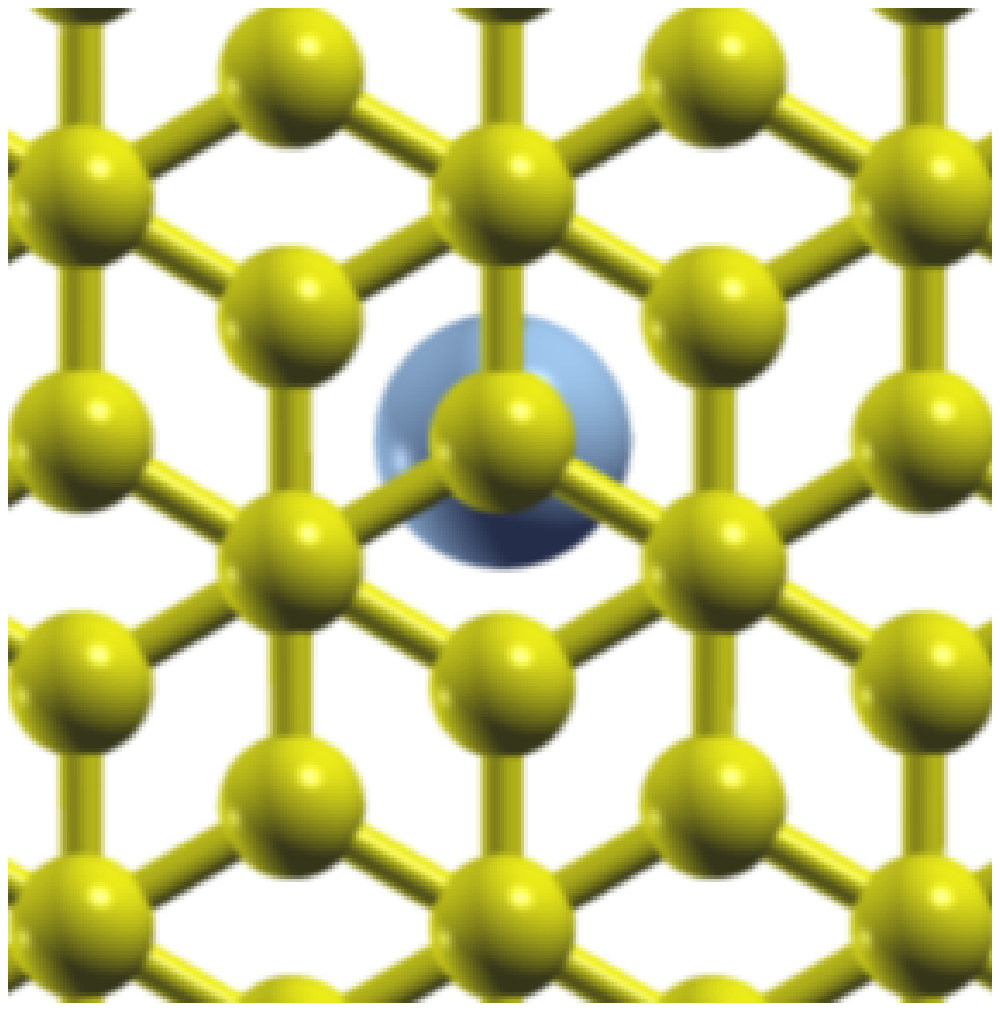}&\includegraphics[scale=0.3]{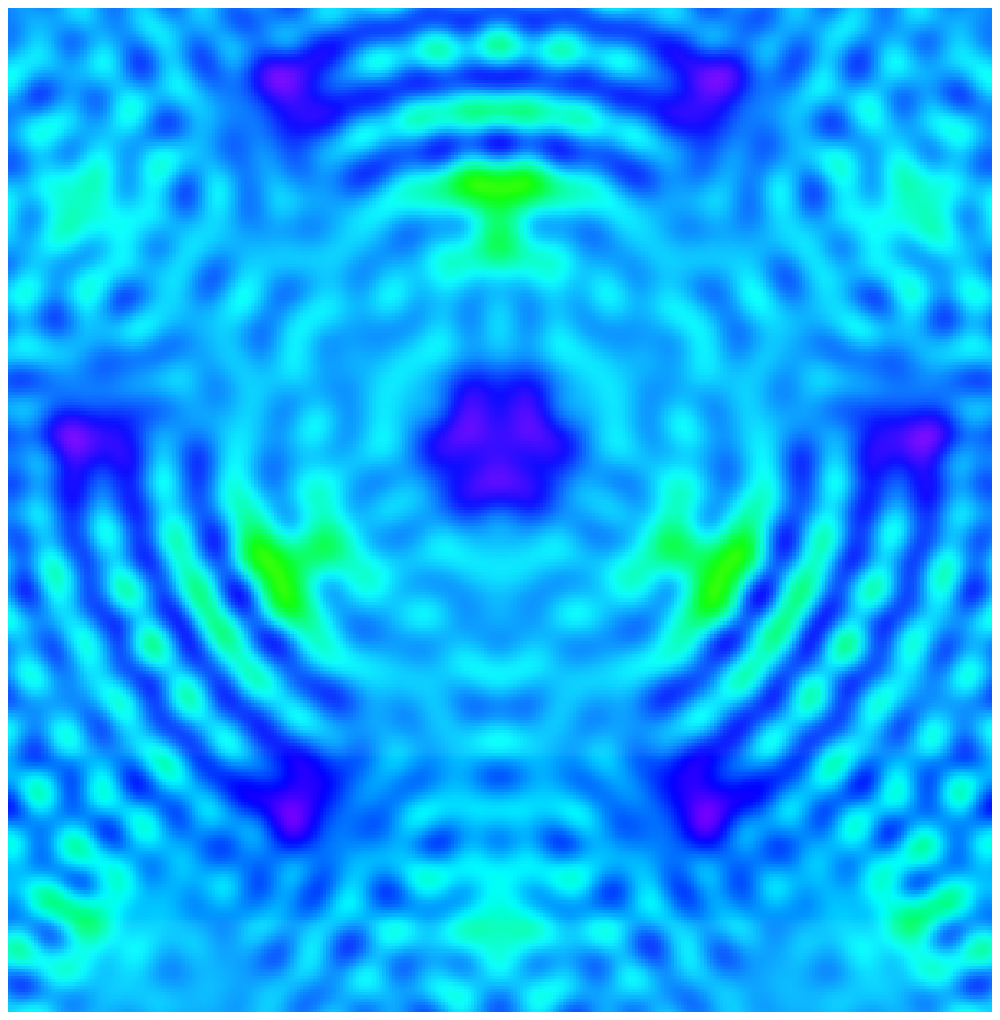}&\includegraphics[scale=0.45]{t}\tabularnewline
\hline 
\end{tabular}\caption{Relaxed structures of the modified graphitic systems 
with intercalated TM atoms in aaa and aba stacking. 
The induced spin densities are compared to each other.}
\end{figure*}

\begin{table*}[p]
\begin{tabular}{|c|c|c|c|c|c|c|c|}
\hline
Stacking  & Dopant & Magnetic moment ($\mu_B$)  &   $l_{TM}$ (\AA)  &   $E_{coh}$ (eV)  &  $E_b$ (eV)  &   $l_C$ (\AA)  &   $l_B$ (\AA)     \\
\hline

    & Mn        &4.10   &2.00   &8.93 &4.22      &1.43$-$1.45 &0.18    \\

    & Fe       &3.70    &2.00   &8.90 &3.69      &1.43$-$1.44 &0.09     \\

aaa & Co       &1.97    &2.02   &8.92 &4.03      &1.43$-$1.44 &0.10     \\

    & Ni       &0.00    &1.99   &8.91 &4.12      &1.43$-$1.44 &0.17     \\

    & Cu       &0.00    &1.98   &8.82 &1.13      &1.43$-$1.45 &0.01     \\

\hline
    & Mn    &   3.80     &2.19   &8.91 &4.29      &1.43$-$1.45 &0.12    \\
                                                                       
    & Fe    &   2.06     &2.17   &8.89 &3.42      &1.43$-$1.45 &0.01    \\
                                                                       
aba & Co    &   1.83     &2.26   &8.92 &4.15      &1.43$-$1.45 &0.08    \\
                                                                       
    & Ni    &   0.00     &2.15   &8.90 &4.02      &1.43$-$1.44 &0.04    \\
                                                                       
    & Cu    &   0.00     &2.04   &8.82 &0.82      &1.43$-$1.45 &0.09    \\
\hline

\end{tabular}
\caption{Magnetic moment (per unit cell), 
average distance between dopant and nearest neighbour C atoms ($l_{TM}$), 
cohesive energy ($E_{coh}$), in-plane C$-$C bond length ($l_C$), bonding energy ($E_b$), 
and buckling of the graphene layers ($l_B$).}
\end{table*}

\begin{figure*}[p]
 \includegraphics[width=15cm,height=15cm,keepaspectratio]{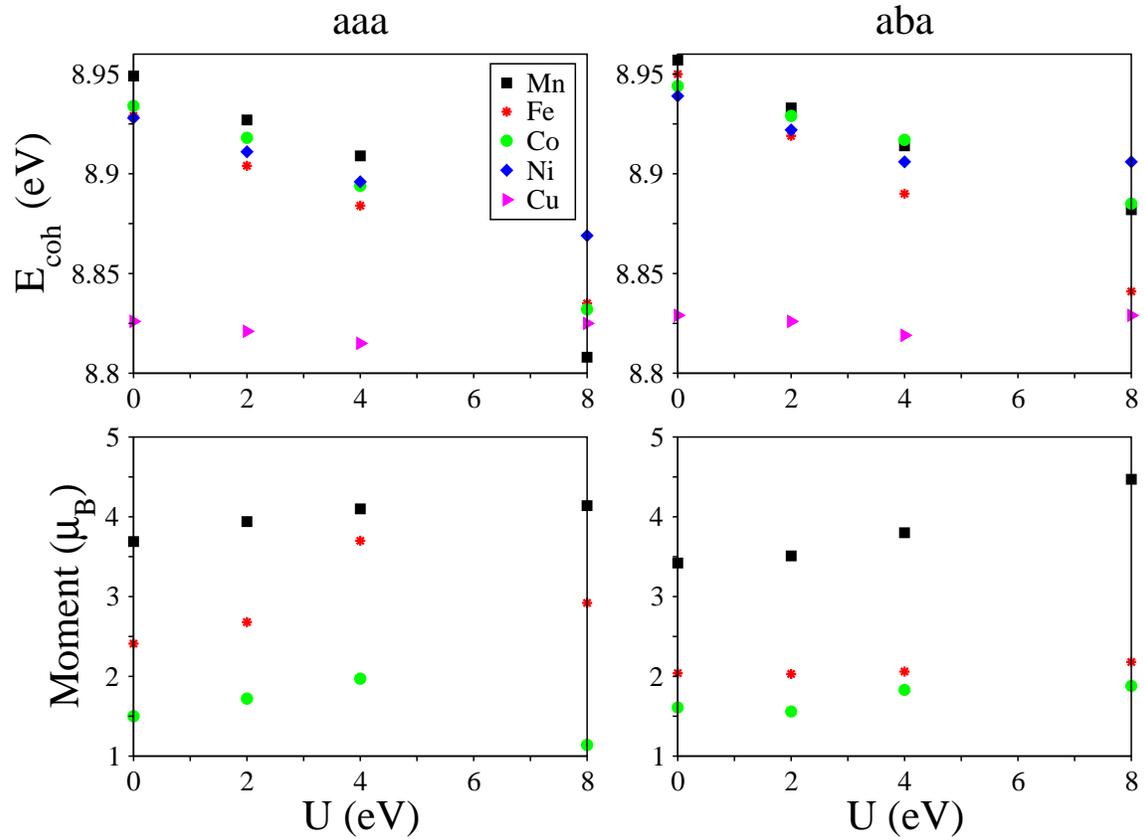}
\caption{Variation of the cohesive energy ($E_{coh}$) and total magnetic moment (per unit cell)
with the onsite energy ($U$) in aaa and aba stacking. }
\end{figure*}

\begin{figure*}[p]
 \includegraphics[width=15cm,height=15cm,keepaspectratio]{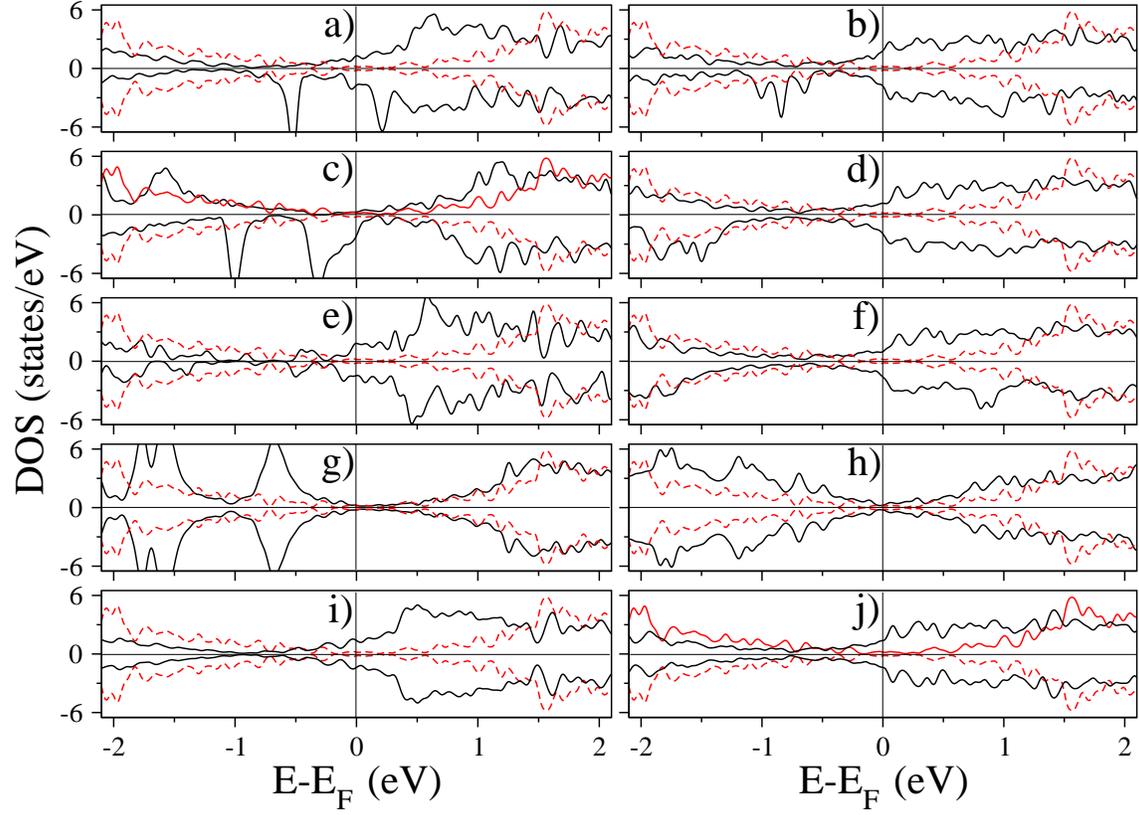}
\caption{Full lines: Spin majority and minority DOS obtained for the configurations (a) Mn(aaa), (b) Mn(aba), (c) Fe(aaa), (d) Fe(aba), 
(e) Co(aaa), (f) Co(aba), (g) Ni(aaa), (h) Ni(aba), (i) Cu(aaa), and (j) Cu(aba), where U = 4 eV. Dashed lines: Corresponding DOS for pristine graphite.}
\end{figure*}
\end{document}